\documentclass{rnaastex}
\usepackage{amsmath,color,textcomp,url,graphicx,subfigure}
\usepackage{morefloats}

\newcommand{\msol}{$M_{\odot}$}
\newcommand{\kms}{\hbox{km\,s$^{-1}$}}

\begin{document}

\title{A PRE-\emph{Gaia}~DR2 SURVEY FOR NEARBY M DWARFS IN YOUNG ASSOCIATIONS}

\author[0000-0002-2592-9612]{Jonathan Gagn\'e}
\affiliation{Carnegie Institution of Washington DTM, 5241 Broad Branch Road NW, Washington, DC~20015, USA}
\affiliation{NASA Sagan Fellow}
\email{jgagne@carnegiescience.edu}
\author[0000-0001-6251-0573]{Jacqueline K. Faherty}
\affiliation{Department of Astrophysics, American Museum of Natural History, Central Park West at 79th St., New York, NY 10024, USA}
\author[0000-0002-2357-1012]{Gilles Fontaine}
\affil{Institute for Research on Exoplanets, Universit\'e de Montr\'eal, D\'epartement de Physique, C.P.~6128 Succ. Centre-ville, Montr\'eal, QC H3C~3J7, Canada}


\section{}

Members of young associations are currently incomplete in the low-mass star regime ($\lesssim$\,0.6\,\msol). Two recent data sets (RECONS and URAT-South; \citealp{2015AJ....149....5W,2018AJ....155..176F}) published trigonometric distances for 2664 low-mass stars, making it possible to identify new candidate members just before data release 2 of the \emph{Gaia} mission \citep{2016AA...595A...1G}. We used the Bayesian classification algorithm BANYAN~$\Sigma$ \citep{2018ApJ...856...23G} to identify new candidate members in 27 young associations from this sample based on their sky position, proper motion, trigonometric distance. We complemented our sample with 2MASS $J$-band magnitudes, \emph{Gaia} $G$-band magnitudes, and literature radial velocities when available. All objects were cross-matched with known bona fide members \citep{2018ApJ...856...23G} and new candidate members from \emph{Gaia}--TGAS \citep{Gagne:2018un}, and the 11 matches were excluded. We selected all targets with a Bayesian young association membership probability above 90\%, and an optimal $UVW$ distance from the BANYAN~$\Sigma$ kinematic models below 5\,\kms\ and 5$\sigma$.

All resulting 146 stars (listed in Table~\ref{table}) were cross-matched with literature catalogs to gather spectral types, signs of youth (e.g., X-ray, UV, Li absorption, infrared excess), and to identify the previously known members or candidate members. Most are M dwarfs, and 131 are newly identified candidate members of young associations. Three of the new candidates (G~13--39 and LHS~2935 in the Carina-Near moving group, and G~152--1 in the AB~Doradus moving group, or ABDMG hereafter) have complete kinematics, but further observations will be needed to confirm their young age determine if they are new bona fide members. We also identified the helium-atmosphere white dwarf EGGR~344 as a candidate member of the 130--200\,Myr-old ABDMG \citep{2015MNRAS.454..593B}, but its effective temperature of $\sim$\,6950\,K \citep{2015ApJS..219...19L} corresponds to a cooling age of $\sim$\,1.3\,Gyr that is incompatible with this hypothesis, even with a conservatively low mass of 0.5\,\msol\ \citep{2001PASP..113..409F}. We therefore reject EGGR~344 as a candidate member of ABDMG.

This catalog of new potentially young M dwarfs, which will soon be improved with \emph{Gaia}~DR2, will be especially useful to characterize the initial mass function of young associations, calculate lithium depletion boundary ages, study their angular momentum evolution, and search for exoplanet companions by direct imaging.

\begin{longrotatetable}
\global\pdfpageattr\expandafter{\the\pdfpageattr/Rotate 90}
\begin{deluxetable*}{lllllr@{\;$\pm$\;}lr@{\;$\pm$\;}lr@{\;$\pm$\;}lcclll}
\tablecolumns{16}
\tablecaption{New candidate members. \label{table}}
\tablehead{\colhead{Name} & \colhead{Assoc.\tablenotemark{a}} & \colhead{Spectral} & \colhead{R.A.} & \colhead{Decl.} & \multicolumn{2}{c}{$\mu_\alpha\cos\delta$}  & \multicolumn{2}{c}{$\mu_\delta$} & \multicolumn{2}{c}{Distance} & \colhead{HR1} & \colhead{$NUV$} & \colhead{IR$_{\rm ex}$} & \colhead{Known} & \colhead{Youth\tablenotemark{c}}\\
\colhead{} & \colhead{} & \colhead{Type\tablenotemark{b}} & \colhead{(hh:mm:ss)} & \colhead{(dd:mm:ss)} & \multicolumn{2}{c}{($\mathrm{mas}\,\mathrm{yr}^{-1}$)} & \multicolumn{2}{c}{($\mathrm{mas}\,\mathrm{yr}^{-1}$)} & \multicolumn{2}{c}{(pc)} & \colhead{} & \colhead{(mag)} & \colhead{$N\sigma$} & \colhead{Group} & \colhead{}
}
\startdata
LP~524--13 & ABDMG & M3 & 00:02:04.017 & +04:08:09.95 & $183.3$ & $3.3$ & $-201.4$ & $3.2$ & $22.3$ & $3.0$ & $\cdots$ & $\cdots$ & $\cdots$ & ABDMG & $\cdots$\\
G~171--61 & ABDMG & M1.5 & 00:32:00.621 & +43:56:33.02 & $296.4$ & $3.9$ & $-268.1$ & $3.9$ & $21.5$ & $2.5$ & $\cdots$ & $\cdots$ & $\cdots$ & $\cdots$ & $\cdots$\\
LP~295--636 & ABDMG & M3 & 01:12:55.574 & +27:44:51.43 & $152.3$ & $4.1$ & $-214.2$ & $4.1$ & $23.7$ & $2.9$ & $\cdots$ & $\cdots$ & $\cdots$ & $\cdots$ & $\cdots$\\
G~244--32 & ABDMG & (M6) & 01:42:27.090 & +61:02:29.49 & $244.8$ & $3.1$ & $-213.2$ & $3.1$ & $23.2$ & $2.2$ & $-0.47$ & $\cdots$ & $\cdots$ & $\cdots$ & $\cdots$\\
UCAC4~485--002908 & ABDMG & (M3) & 02:03:25.984 & +06:47:59.15 & $89.2$ & $1.2$ & $-120.5$ & $1.2$ & $23.5$ & $2.8$ & $\cdots$ & $\cdots$ & $\cdots$ & ABDMG & $\cdots$\\
\enddata
\tablenotetext{a}{Young association. See \cite{2018ApJ...856...23G} for more detail.}
\tablenotetext{b}{Spectral types in parentheses were estimated using the $G - J$ color, similarly to \citep{Gagne:2018un}.}
\tablenotetext{c}{Signs of youth compiled from the literature, see \cite{Gagne:2018un} for more detail.}
\tablecomments{This is a preview of the full electronic-only table.}
\end{deluxetable*}
\end{longrotatetable}
\global\pdfpageattr\expandafter{\the\pdfpageattr/Rotate 0}

\acknowledgments

This research made use of: data products from the Two Micron All Sky Survey, which is a joint project of the University of Massachusetts and the Infrared Processing and Analysis Center/California Institute of Technology, funded by NASA and the National Science Foundation, and data from the ESA mission {\it Gaia}, processed by the {\it Gaia} Data Processing and Analysis Consortium whose funding has been provided by national institutions, in particular the institutions participating in the {\it Gaia} Multilateral Agreement.

\software{BANYAN~$\Sigma$ \citep{2018ApJ...856...23G}.}

\bibliographystyle{apj}

\end{document}